\documentclass[aps,prb,twocolumn,groupedaddress,showpacs,superscriptaddress,amssymb,amsmath,noeprint]{revtex4-2}
\usepackage{graphicx}% Include figure files
\usepackage{dcolumn}% Align table columns on decimal point
\usepackage{bm}% bold math
\usepackage{hyperref}% add hypertext capabilities
\usepackage{cleveref}
\hypersetup{
    colorlinks=true,
    linkcolor=blue,
    urlcolor=blue,
	citecolor=blue
}
\usepackage{comment}
\usepackage{color}
\usepackage[utf8]{inputenc}
\usepackage{graphicx}
\usepackage{tabularx}
\usepackage{xcolor}
\usepackage{amsmath}
\usepackage{dcolumn}
\usepackage{hyperref}
\usepackage{bm}
\usepackage{epsf}
\usepackage{braket}
\usepackage{tensor}
\usepackage{soul}

\begin{document}
\newcolumntype{M}[1]{>{\centering\arraybackslash}m{#1}}
%\preprint{APS/123-QED}

\title{Assessment of spectral phases of non-Hermitian quantum systems through complex and singular values}

\author{Mahaveer Prasad}
\email{mahaveer.prasad@icts.res.in} 
\affiliation{International Centre for Theoretical Sciences, Tata Institute of Fundamental Research,
Bangalore 560089, India}

\author{S. Harshini Tekur}
\email{saiharshini.t@prayoga.org.in} 
\affiliation{Prayoga Institute of Education Research, Bangalore 560116, India}

\author{Bijay Kumar Agarwalla}
\email{bijay@iiserpune.ac.in}
\affiliation{Department of Physics, Indian Institute of Science Education and Research, Pune 411008, India}

\author{Manas Kulkarni}
\email{manas.kulkarni@icts.res.in} 
\affiliation{International Centre for Theoretical Sciences, Tata Institute of Fundamental Research,
Bangalore 560089, India}

\date{\today} 

\begin{abstract}
Chaotic behavior or lack thereof in non-Hermitian systems is often diagnosed via spectral analysis of associated complex eigenvalues. Very recently, singular values of the associated non-Hermitian systems have been proposed as an effective measure to study dissipative quantum chaos. Motivated by the rich properties of non-Hermitian power-law banded random matrices and its promise as a platform to study localized and delocalized phases in non-Hermitian systems, we make an in-depth study to assess different spectral phases of these matrices through the lens of both complex eigenvalues and singular values. Remarkably, the results from complex spectra and singular value analysis are seemingly different, thereby necessitating caution while identifying different phases. We also exemplify our findings by studying a non-Hermitian Hamiltonian with a complex on-site disorder. Our work indicates that systems, where disorder is present both in the Hermitian and non-Hermitian segments of a Hamiltonian, are sensitive to the specific diagnostic tool that needs to be employed to study quantum chaos.

\end{abstract}

\maketitle

{\it Introduction.--}
Chaos in quantum systems is usually diagnosed by studying the spectral statistics of the appropriate Hamiltonian \cite{IZRAILEV1990299, casati_chaos, Haake_2001, Reichl2004,gutzwiller2013chaos}. The correspondence to Random Matrix Theory (RMT) \cite{mehta2004random, Dyson_1962_1, Dyson_1962_2, GUHR1998189, beenakker1997random} is well-established via the Bohigas-Giannoni-Schmidt conjecture \cite{BGSconjecture} and the Berry-Tabor conjecture \cite{BerryTaborconjecture} in the chaotic and integrable/regular limits, respectively. The quantities associated with the statistics of eigenvalues and eigenvectors, namely, the level spacing distribution \cite{Wigner_1951, Porter_1965, PhysRevE.73.036201, Srivastava_2019, PhysRevB.102.054202}, distribution of spacing ratios \cite{PhysRevLett.110.084101, Atas_2013, PhysRevE.101.022222, Harshini_higher_order}, spectral form factor \cite{PhysRevE.55.4067, PhysRevLett.121.264101,PhysRevResearch.3.L012019,Prasad_2024, PhysRevE.109.064208, PhysRevLett.134.010402, das2025spectralformfactorenergy}, inverse participation ratio \cite{PhysRevLett.71.412, PhysRevB.98.054206, PhysRevLett.84.3690} and other statistics \cite{GUHR1998189,PhysRevLett.100.044103, Rigol2016, PhysRevX.10.041017} are robust and well-understood for Hermitian quantum systems in both limits and even in the mixed and transition regime from integrability to chaos \cite{brody1973statistical, MVBerry_1984, Seligman1984, Seligman_1985, Izrailev_1989, Prosen_1993, Prosen_1994, PhysRevE.97.062212}. 

However, the same level of understanding has not been achieved for dissipative quantum chaotic systems, which are often described either by the Liouvillian corresponding to a Markovian Lindblad master equation \cite{OQS_Breuer, rotter2015review}, or by an effective non-Hermitian Hamiltonian (NHH) \cite{bender2007making, RevModPhys.93.015005}. This is primarily because the corresponding eigenvalues are complex, and consequently not as easy to analyze compared to its Hermitian counterpart. Some aspects of non-Hermitian systems have been extensively studied, for example, the classification of the symmetry classes of NHHs \cite{PhysRevB.55.1142, Bernard2002, PhysRevResearch.2.023286, NH_Jacobus, PhysRevX.13.031019, PRXQuantum.4.030328, PhysRevResearch.5.033043}, the observation of localized-delocalized transitions \cite{PhysRevLett.77.570, PhysRevLett.102.065703, PhysRevB.98.020202, NHMBLHKU2019, Scipost_Dukelsky},  characterization using RMT measures \cite{jung1999phase, buijsman2019, Jaiswal_2019, 10.21468, Can_2019, Sa_2020, PhysRevLett.123.140403, PRR_Ryuichi, yang2024decoherence, xiao2024universal, PhysRevLett.124.100604, Chenu2025}, as well as the correspondence between dissipative quantum chaotic systems and level repulsion in non-Hermitian random matrices \cite{PhysRevE.55.205, prosen_PRL_2019, PhysRevA.105.L050201, Lange_Timm_2021, Santos2024} (the so-called Grobe-Haake-Sommers conjecture \cite{Haake_regular_chaotic, GrobePRL1989}). Quantities such as complex spacing ratios (CSRs) \cite{complex_prosen, Yusipov_chaos} and dissipative spectral form factors \cite{PhysRevLett.127.170602, Yoshimura_Sa_2024, GGK2022}, have become popular and powerful diagnostic tools to analyze short-range and long-range correlations, respectively. Very recently, the statistics of singular values \cite{svd_Zhenyu, chenu_svd, PhysRevE.110.064307_svd, nandy2024krylovspaceapproachsingular, Tekur_svd, nandy_svd} has proved to be an alternate promising avenue into furthering our understanding of quantum chaos for dissipative quantum systems. Since the spectral analysis in this case involves only real numbers, this makes it not only computationally amenable but also more straightforward to apply the traditional RMT diagnostics.

Despite these two complementary promising approaches, it remains unclear to what extent the correlations present in complex eigenvalues are encoded in the  singular values and whether the corresponding predictions are comparable. Particularly, it is not obvious whether these quantities capture similar behavior, especially in the regions that show transitions between different phases. In this Letter, we perform a careful analysis to understand the nature of the transition as captured by these different diagnostics, for the non-Hermitian version of a well-known class of random matrices, namely the Power-Law Banded Random Matrices \cite{PhysRevE.54.3221, PhysRevB.61.R11859, PhysRevB.74.125114, PhysRevB.82.125106, PhysRevE.98.042116, Rao_2022,buijsman2025powerlawbandedrandommatrix}. This setup has recently gained significant attention as it is well-suited for studying delocalization-localization transitions in disordered, non-Hermitian systems \cite{PhysRevB.108.L180202, PhysRevB.108.L060201}. We next consider a class of physical NHHs (i) that contains disorder in both its real and imaginary parts and (ii) that contains disorder solely in the imaginary (non-Hermitian) part and study the predictions that follow from the complex eigenvalue and singular value statistics.

In what follows, we describe the diagnostic tools that we use for our analysis. The adjacent gap ratio is a popular measure of level repulsion which is defined for a real spectrum $\{ E \}$ as $r_i = \frac{\mathrm{min}(\delta_i,\delta_{i+1})}{\mathrm{max}(\delta_i,\delta_{i+1})}$, where $\delta_i =E_{i+1} - E_i$ is the nearest level eigenvalue spacing computed from the spectrum. This quantity has been extended to the non-Hermitian systems characterized by complex eigenvalue spectra ($\{ z \}$) \cite{complex_prosen} and is defined as
    \begin{equation}
    \label{eq:complex_r}
    z_i = \frac{E_i-E_{\rm NN}}{E_i-E_{\rm NNN}}= r_i e^{i \theta_i},
    \end{equation}
    where $E_{\rm NN}$   $(E_{\rm NNN})$ is the nearest neighbor (next-nearest neighbor) of the complex eigenvalue $E_i$. Very recently, the singular values (eigenvalues of $\sqrt{M^{\dagger} M}$, where $M$ is the non-Hermitian matrix representing the system) associated with the non-Hermitian systems have emerged as a promising route to characterize chaotic properties \cite{svd_Zhenyu, chenu_svd, PhysRevE.110.064307_svd, nandy2024krylovspaceapproachsingular, Tekur_svd, nandy_svd}. Since the singular values are real and non-zero, the gap ratio is computed using the standard definition employed for Hermitian systems. The average of the computed ratios, which we generically denote as $\langle r\rangle$, provides an estimate of whether the system is in the chaotic or localized limit or the transition region. We denote this quantity as $\langle r\rangle_{\rm CSR}$ and $\langle r\rangle_{\rm SV}$ for ratios computed using complex eigenvalues and singular values, respectively, with $\langle r\rangle_{\rm CSR} = \langle |z|\rangle$. We also compute the singular form factor $\sigma$FF~\cite{chenu_svd, nandy_svd}, an extension of the spectral form factor (SFF) to singular values, which can be calculated for a non-Hermitian matrix $M$ as,
\begin{equation}
    \sigma FF (t)= \Bigg| \frac{1}{\cal N} \sum _{n=1}^{\cal N} e^{-i \sigma_n t} \Bigg|^2 =|\langle \psi_R|e^{-i \sqrt{M^{\dagger} M}t}|\psi_R \rangle|^2,
    \label{eq:sigma_FF}
\end{equation}
where ${\cal N}$ is the dimension of Hilbert space, $\sigma_n$ are the singular values of $M$ (obtained after following the standard unfolding procedure) and $|\psi_R\rangle= \sum_{n=1}^{\cal N} |v_n\rangle/\sqrt{\cal N}$ is the right infinite-temperature coherent Gibbs state with $|v_n\rangle$ being the right singular vector of $M$ (eigenvector of the matrix $M^{\dagger} M$) \cite{chenu_svd}. When $M$ belongs to the  non-Hermitian Ginibre  unitary random matrix class, then its singular values follow the Gaussian Unitary Ensemble (GUE)-RMT class, and in this case, the singular form factor is given as \cite{PhysRevD.98.086026}  
\begin{equation}
\sigma FF_{\rm GUE}(t)=
\begin{cases}
        t/t_{H},   \ \  t \leq t_H \\
        1, \ \  \  t \geq t_H
\end{cases}
\label{sigmaFF_GUE}
\end{equation}
where $t_H$ is the Heisenberg time. On the other hand, using complex eigenvalues, the so-called dissipative spectral form factor (DSSF) can be computed for non-Hermitian systems, which is defined as \cite{DSFF_2021} 
\begin{equation}
K(\tau,\tau^{*})= \Bigg| \frac{1}{\cal N} \sum_{n=1}^{\cal N} e^{i (z_n \tau^{*} + z_n^{*} \tau)/2}\Bigg|^2.
\label{DSFF}
\end{equation}
DSFF for Poissonian random spectrum \cite{DSFF_2021} that is corresponding to integrable systems decay exponentially as ${K(\tau,\tau^{*})}_{\rm Poisson}={\cal N}+{\cal N}({\cal N}-1)\,e^{-|\tau|^2}$. In our analysis, we do not perform complex spectra unfolding procedure to compute CSR and DSFF. Note that the CSR is insensitive to the unfolding procedure and essential features of DSFF are 
captured irrespective of whether one employs the unfolding procedure. For DSFF, we project the complex eigenvalues on the imaginary axis of the complex plane. Finally, we compute the inverse participation ratio (IPR), which is a useful measure to quantify the degree of localization of eigenstates in a quantum system. For a normalized eigenstate $|\psi\rangle $ with components $ |\psi_i \rangle$ in a chosen basis we compute the $\rm IPR^q$ and extract the fractal dimensions $D_f^{q}$ as follows \cite{PhysRevB.108.L180202},
\begin{equation}
    {\rm IPR^{q}}=\sum_{i=1}^{\cal N} |\psi _i|^{2q}, \ \ \ \ \ \ D_f^{q}=\frac{1}{1-q}\frac{d \, \rm ln \, IPR^{q}}{d\,\rm ln \, {\cal N}}. 
    \label{IPR-frac}
\end{equation}
For a non-Hermitian system with complex spectra, we compute $\rm IPR^{q}$  using the right eigenvectors with $q=2$. Note that for a strongly localized state, ${\rm IPR} \approx 1$, and for a completely delocalized (extended) state, ${\rm IPR}=\frac{1}{\cal N}$, eventually vanishing in the thermodynamical limit. For the singular values, we compute $\rm IPR^{q}$ and fractal dimension using the right singular vectors. In what follows, we perform a detailed analysis of the nature of the transition as captured by these different diagnostics. In our analysis we consider the entire spectrum of the associated non-Hermitian setup. We first focus on the non-Hermitian version of power-law banded random matrices.

\begin{figure}[h]
    \centering
    \includegraphics[width=0.47\linewidth]{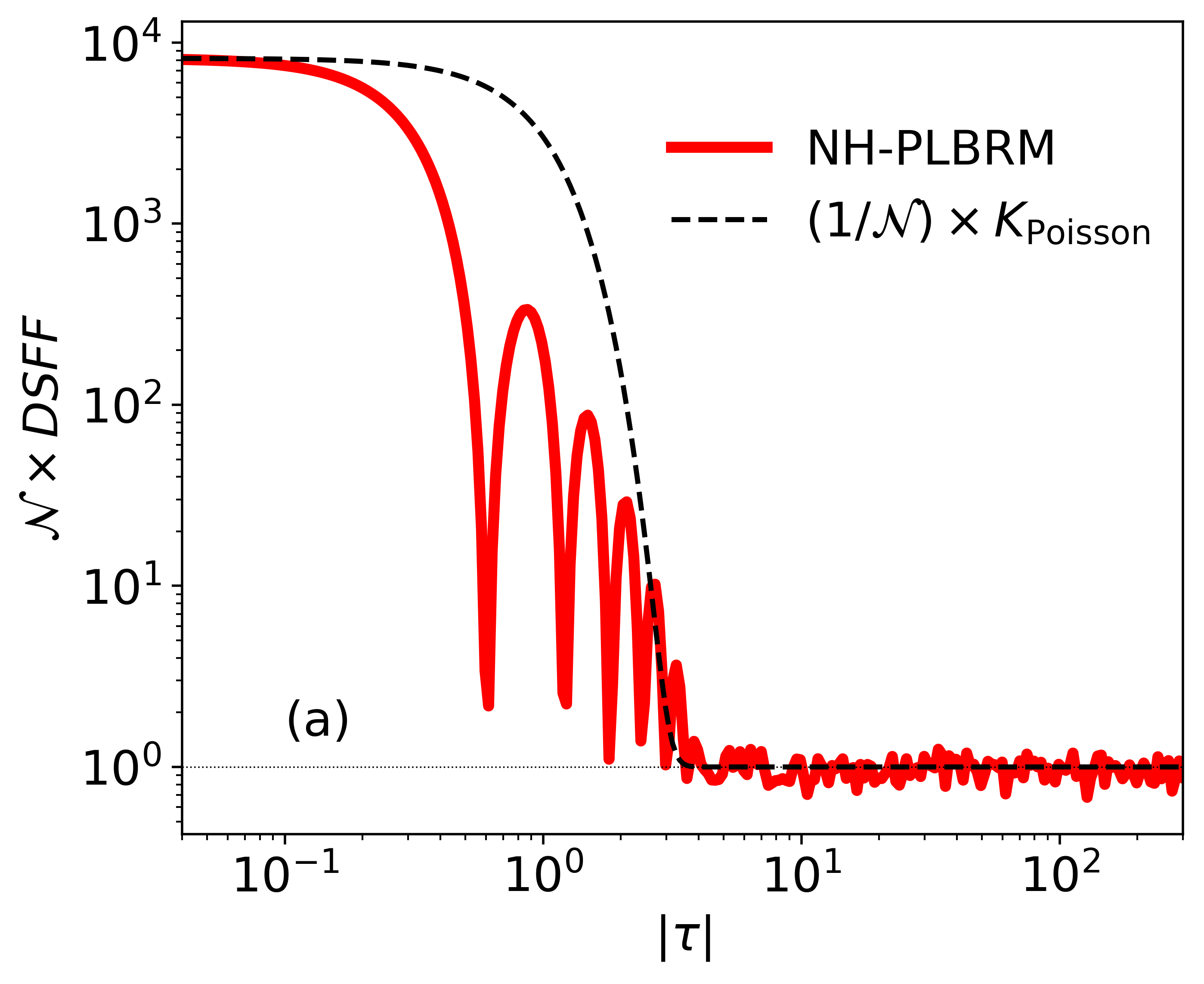}
    \includegraphics[width=0.47\linewidth]{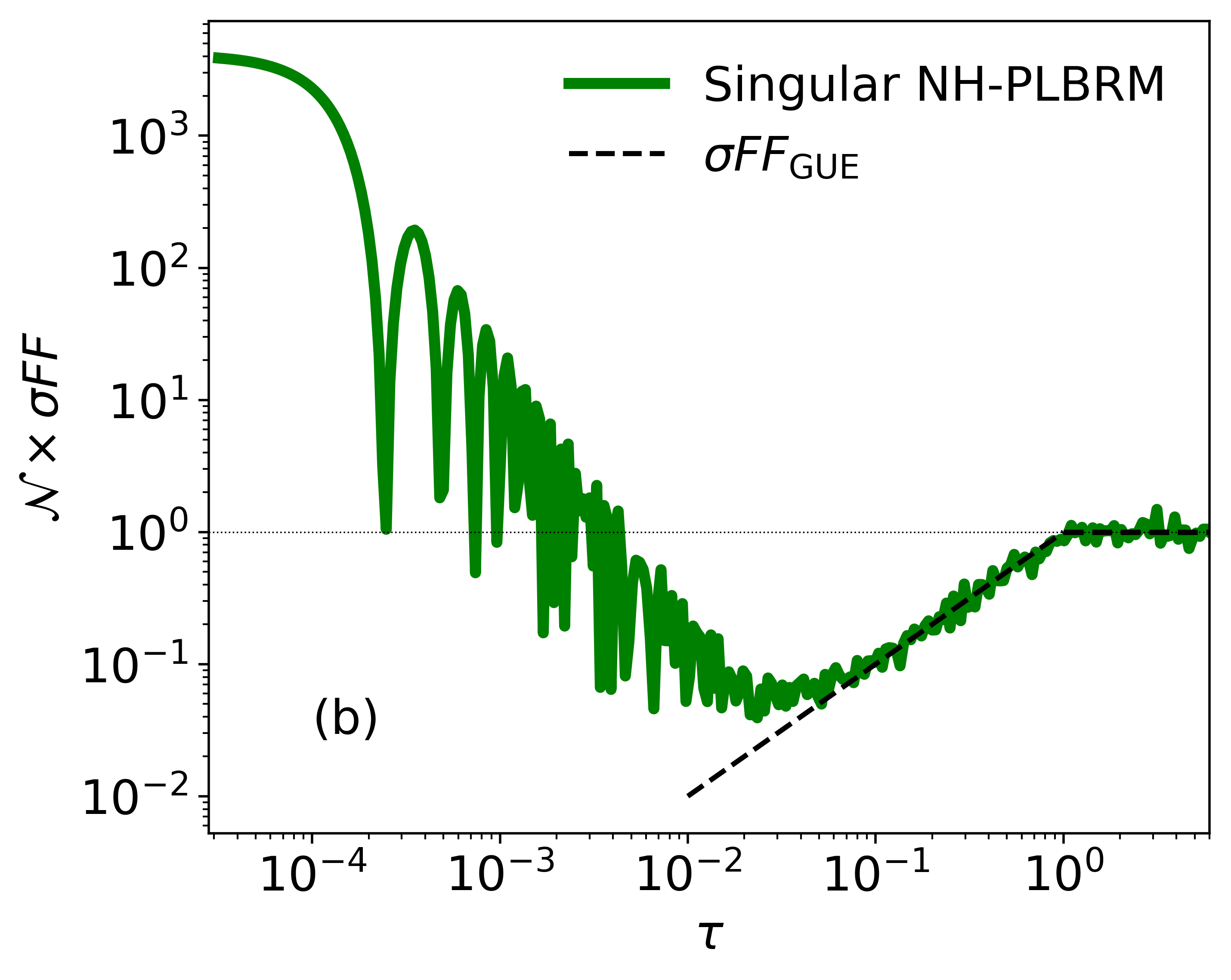}
    \caption{Plot for the (a) dissipative spectral form factor, given in Eq.~\eqref{DSFF} and (b) singular form factor, given in Eq.~\eqref{eq:sigma_FF} for the NH-PLBRM model for $W=5$ and $\alpha=0.8$. The matrix size is chosen to be ${\cal N}=8192$, and averaged over $60$ realizations. The dashed line in (a) represents the analytical prediction that follows from the two-dimensional Poisson distribution ${K(\tau,\tau^{*})}_{\rm Poisson}={\cal N}+{\cal N}({\cal N}-1)\,e^{-|\tau|^2}$. The dashed line in (b) represents the analytical prediction that follows from the spectral form factor for GUE random matrices and is given in Eq.~\eqref{sigmaFF_GUE}.}
    \label{fig:enter-label}
\end{figure}

{{\it Non-Hermitian power-law banded random matrix} (NH-PLBRM).--} We here study a rich class of non-Hermitian random matrix models, known as the non-Hermitian power-law banded random matrix (NH-PLBRM), that has been of significant recent focus \cite{PhysRevE.54.3221, PhysRevB.61.R11859, PhysRevB.74.125114, PhysRevB.82.125106, PhysRevE.98.042116, Rao_2022,buijsman2025powerlawbandedrandommatrix}. The Hermitian analog of NH-PLBRM has been studied extensively and serves as a paradigmatic example for delocalization-localization transitions in quantum systems \cite{PhysRevLett.64.547,Levitov_1989,PhysRevE.54.3221,PhysRevB.62.7920,PhysRevLett.84.3690}. Interestingly, the non-Hermitian version is also known to exhibit delocalization-localization transition \cite{PhysRevB.108.L180202}. Moreover, the non-Hermitian analog is shown to exhibit localization in a parameter window where it ceases to exist in the Hermitian counterpart. We assess the existence of these different spectral phases using different diagnostic tools constructed out of singular values of the non-Hermitian matrix and make an in-depth comparison with the corresponding diagnostic tools proposed for complex eigenvalue analysis. We write the NH-PLBRM as \cite{PhysRevB.108.L180202},
\begin{equation}
\label{eq:power_banded}
    M_{mn}=\epsilon_n \delta_{mn}+J_{mn},
\end{equation}
where $J$ is a Hermitian matrix i.e., $J_{mn}=J_{nm}^{*}$, $\epsilon_n$ and $J_{mn}$ are independent random complex values drawn from a box random distribution as, 
\begin{align}
    |&{\rm Re} (\epsilon_n)|,\  |{\rm Im} (\epsilon_n)| \leq W, \nonumber \\
    |& {\rm Re} (J_{mn})|^2, \ |{\rm Im} (J_{mn})|^2 \leq \frac{1}{2(|m-n|^2 +b^2)^{\alpha}},
\end{align}
where $W$ is the strength of the on-site disorder, $b$ is the band-width of the decay, and $\alpha$ is the exponent associated with the off-diagonal power-law decay.
\begin{figure}[h]
    \centering
    \includegraphics[width=0.49\linewidth]{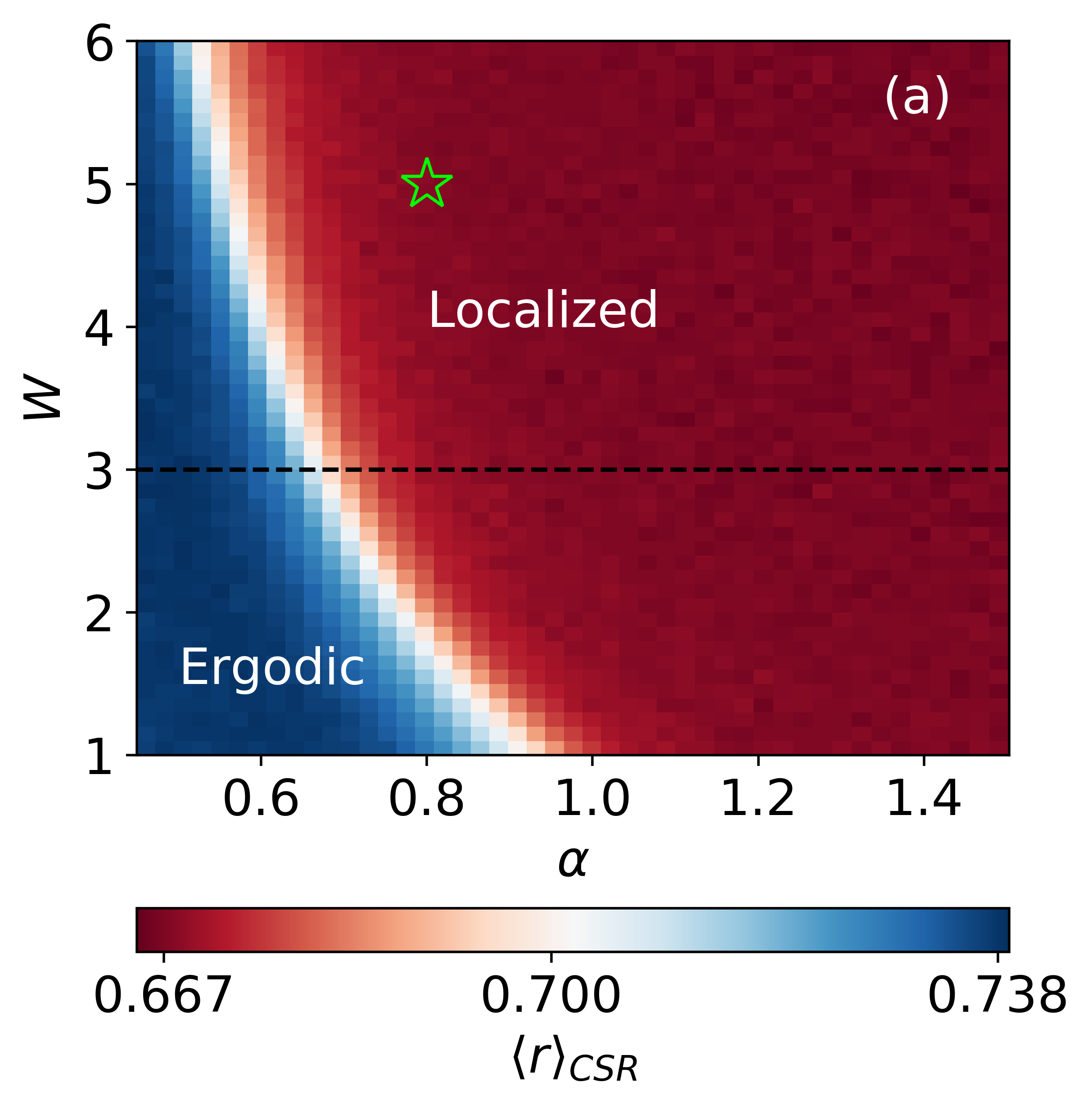}
    \includegraphics[width=0.49\linewidth]{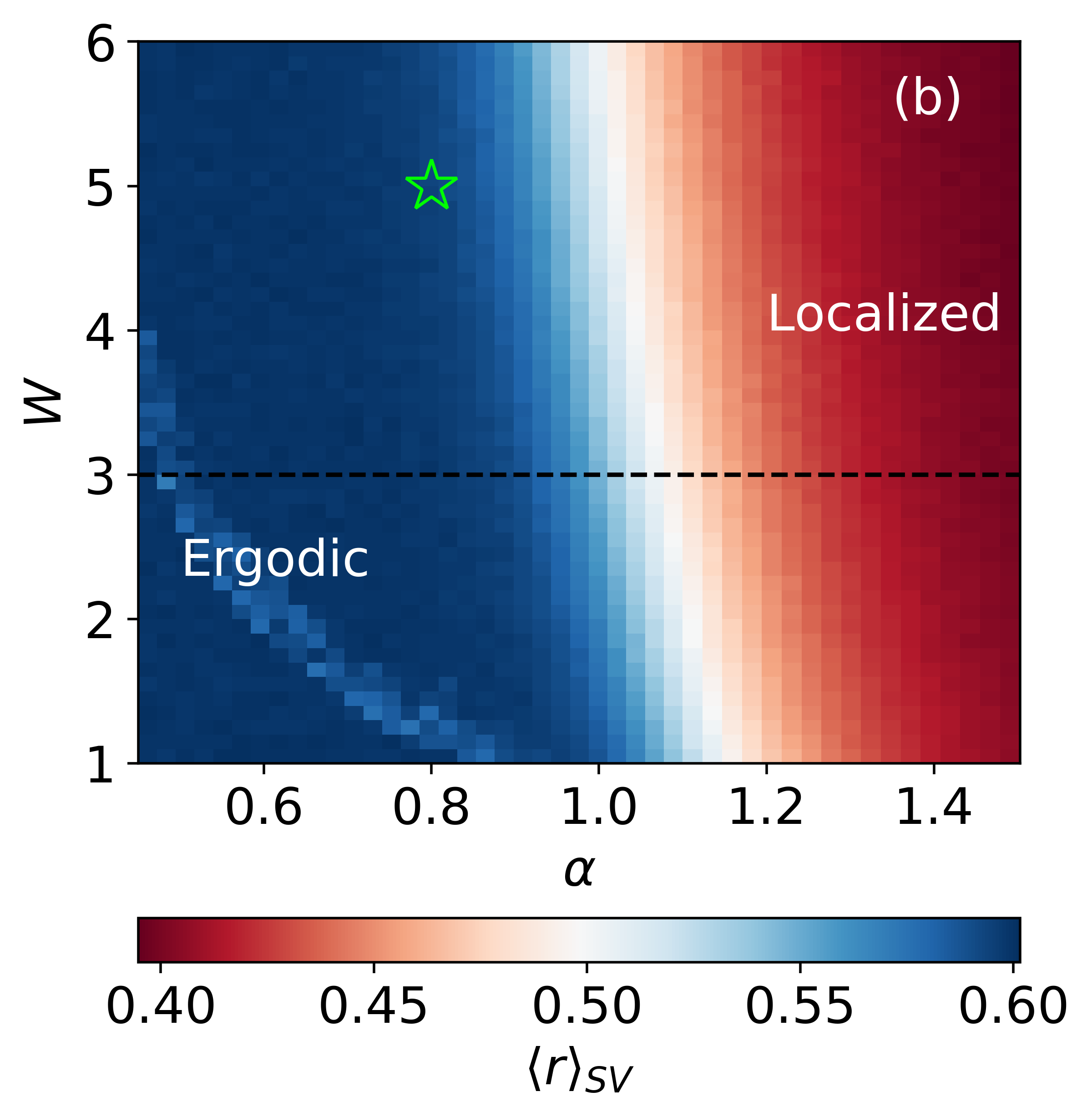}
    \caption{The phase diagram of the NH-PLBRM model, defined in Eq.~\eqref{eq:power_banded}, in the $W\!-\!\alpha$ plane constructed by (a) performing complex spectra analysis and thereby computing $\langle r \rangle_{\rm CSR}$ and (b) performing singular-value statistics and computing $\langle r \rangle_{\rm SV}$. The matrix size is chosen to be ${\cal N}=8192$, and we take 40 different realizations. We take $b=1$ here. The star symbol represents the parameter values that was used to generate dissipative and singular form factors in Fig.~\eqref{fig:enter-label}. The horizontal dashed line represents the range of $\alpha$ values for which inverse participation ratio and fractal dimension are computed in Fig.~\eqref{fig:PLBRM_ipr}.}
    \label{fig:PLBRM_phase_diagram}
\end{figure}
We study the spectral statistics of the NH-PLBRM setup defined in Eq.~\eqref{eq:power_banded} and analyze the phase diagram in the $W\!-\!\alpha$ parameter space. This setup has both the ergodic and the localized regimes, and the transition between these two regions depends on both $\alpha$ and $W$. To this end, we first compute the complex spacing ratio, as given in Eq.~\eqref{eq:complex_r}, using the complex eigenvalues and compute  $\langle r \rangle_{\rm CSR}$. We then compute the adjacent gap ratio of a sequence of real (non-negative) numbers built out of singular values of $M$ in Eq.~\eqref{eq:power_banded} and calculate $\langle r \rangle_{\rm SV}$. In Fig.~\ref{fig:PLBRM_phase_diagram}(a) and (b), we plot the $\langle r \rangle_{\rm CSR}$ and $\langle r \rangle_{\rm SV}$ in the $W\!-\!\alpha$ parameter space, respectively. We notice that both the quantifiers predict the ergodic and localized phases. However, interestingly, the critical line separating these two phases in the $W- \alpha$ plane differs significantly. In particular, the critical line emerging from singular value analysis is markedly similar to the vertical critical line obtained in the Hermitian version of the PLBRM model \cite{Fyodorov_2009}.  Therefore, our analysis reveals that one needs to be cautious when using singular values to diagnose spectral phases. Furthermore, it is natural to study long-range correlations in different phases using both the complex eigenvalues and the singular values. We, therefore, study the dissipative form factor and singular form factor for the complex eigenvalues and singular values, respectively. We pick a representative point that corresponds to a localized phase if one employs CSR as the diagnostic and ergodic if one employs $\langle r\rangle_{\rm SV}$ as the diagnostic [see Fig.~\eqref{fig:PLBRM_phase_diagram}]. It is not immediately obvious that similar differences would be reflected in the form factors. In Fig.~\ref{fig:enter-label}, we provide compelling evidence that the DSFF does not show any sign of a ramp. This absence of a ramp implies that the system is non-ergodic. However, interestingly, the long-range correlations contained in the singular values remarkably show a ramp, thereby indicating ergodic behaviour. As a result, similar to $\langle r \rangle_{\rm CSR}$ and $\langle r \rangle_{\rm SV}$, the form factors also predict completely different spectral phases for the same parameter values of the model.  

\begin{figure}[h]
    \centering
    \includegraphics[width=0.47\linewidth]{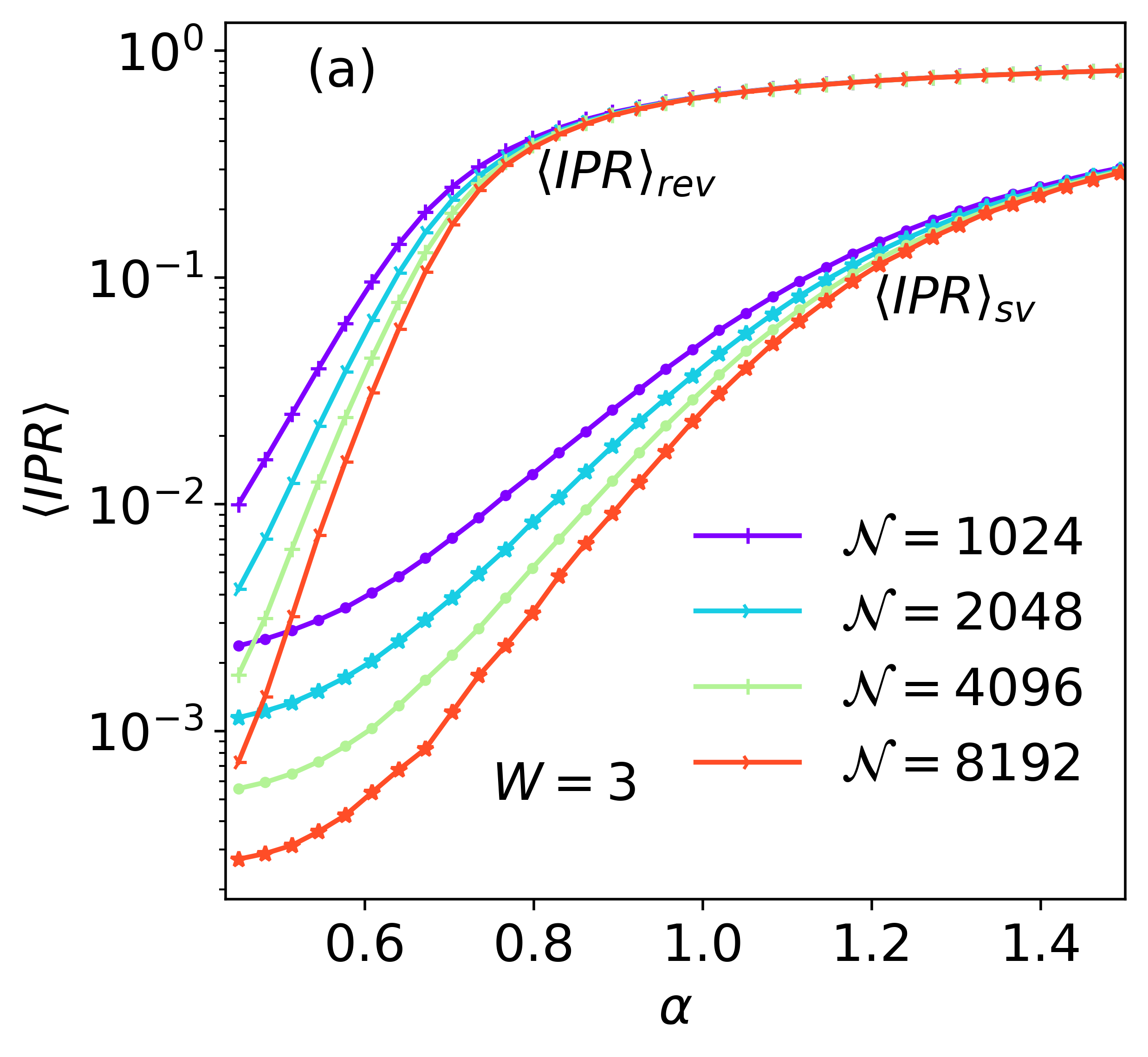}
    \includegraphics[width=0.47\linewidth]{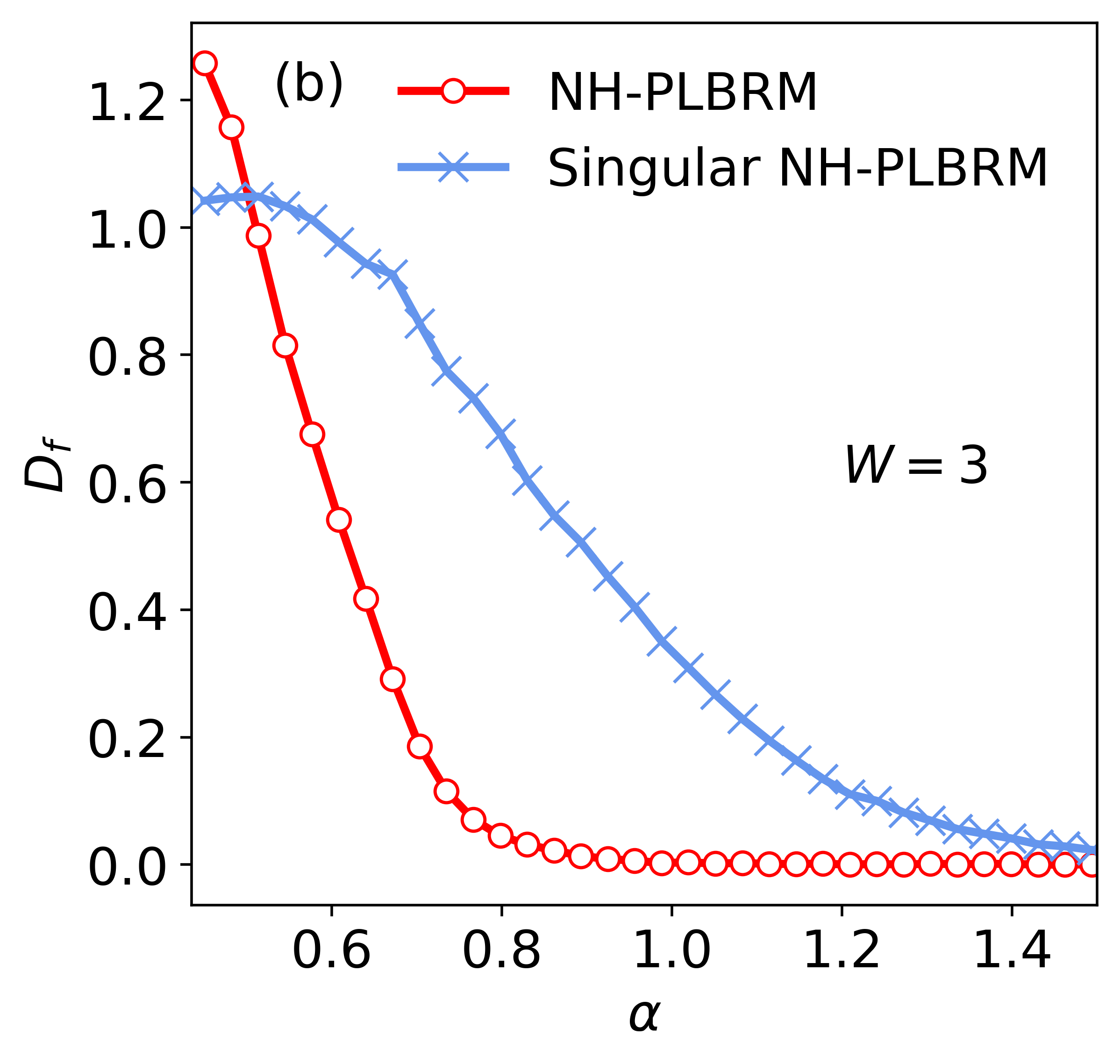}
    \caption{Plots for the (a) inverse participation ratio ($\rm IPR$), defined in Eq.~\eqref{IPR-frac} for $q=2$, constructed following the right eigenvectors of the complex eigenvalues $\langle \rm IPR \rangle_{\rm rev}$ and the IPR calculated using the singular vectors $\langle \rm IPR \rangle_{\rm sv}$ and (b) the fractal dimension $D_f$ for $q=2$  as a function of $\alpha$ for a fixed value of $W=3$ for the NH-PLBRM model. The data is averaged over $300$, $120$, $60$, and $20$ realizations for the matrix size ${\cal N}=1024$, $2048$, $4096$, and $8192$, respectively.}
    \label{fig:PLBRM_ipr}
\end{figure}

Having studied the diagnostics constructed following the eigenvalues, it is essential to look at the predictions that follow by analyzing the eigenvectors corresponding to the complex eigenvalues and the singular vectors corresponding to the singular values. In Fig.~\ref{fig:PLBRM_ipr}, we plot the inverse participation ratio (IPR) and the corresponding fractal dimension, as defined in Eq.~\eqref{IPR-frac} for $q=2$. As clearly seen from Fig.~\ref{fig:PLBRM_ipr}(a), the IPR calculated using the right eigenvector of the complex eigenvalues $\langle \rm IPR \rangle_{\rm rev}$ and the IPR calculated using the singular vectors $\langle \rm IPR \rangle_{\rm sv}$ display completely different transition points as a function of $\alpha$ for fixed $W$. This is also evident from Fig.~\ref{fig:PLBRM_ipr}(b) where the fractal dimension $D_f$ approaches zero at completely different $\alpha$ values. It is worth re-emphasizing that the NH-PLBRM  is an archetypal model for envisaging transitions in non-Hermitian systems. Our findings have clearly demonstrated that the actual spectral transitions seen via complex spectra can potentially be masked when one characterizes the phases merely based on singular values and vectors. 


\begin{figure}[h]
    \centering
    \includegraphics[width=0.47\linewidth]{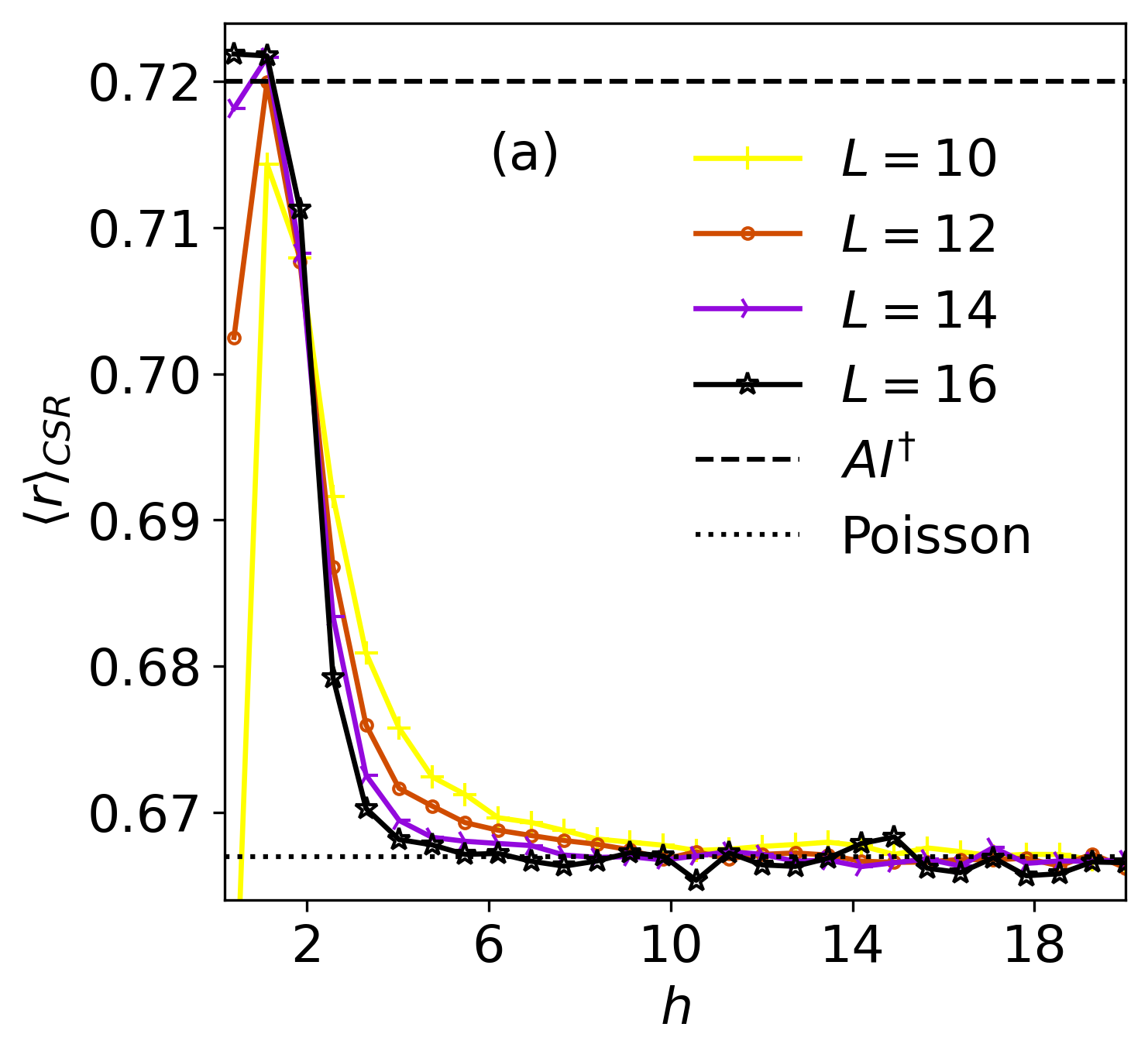}
    \includegraphics[width=0.47\linewidth]{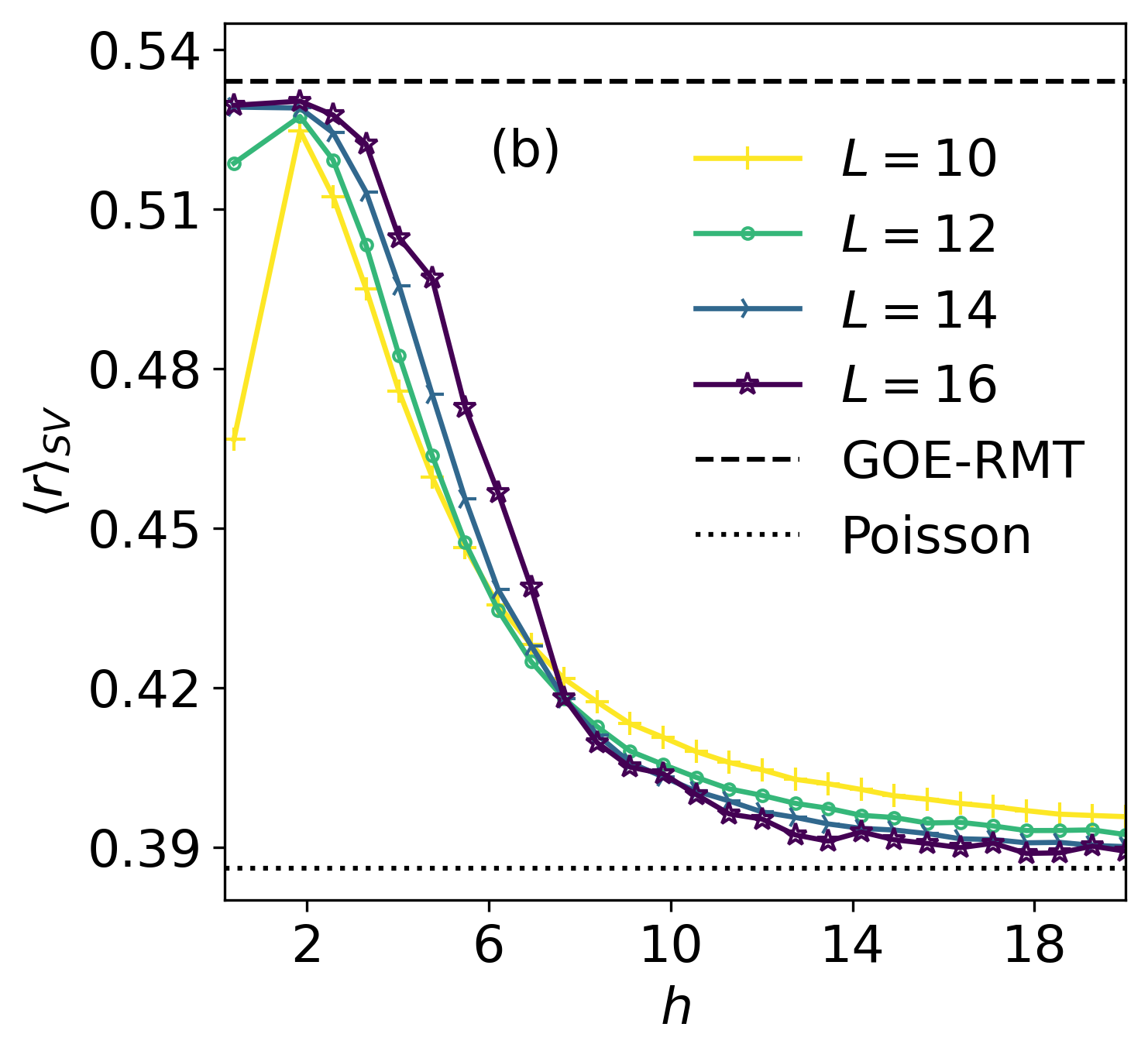}
    \caption{Plots for the average adjacent gap ratio statistics for (a) complex spectra $\langle r \rangle_{\rm CSR}$ and, (b) singular values $\langle r \rangle _{\rm SV}$ of the NHH, defined in Eq.~\eqref{eq:model2} for different system sizes. The top and bottom horizontal lines in (a) and (b) denote the expected values for the corresponding random matrix class ($AI^\dag$ class for CSR and GOE class for SV) and Poisson distributions, respectively. The data is averaged over $12000$, $4000$, $1000$ and $20$ samples for the system size $L=10$, $12$, $14$, and $16$, respectively.}
    \label{fig:model2_r}
\end{figure}

\begin{figure}[h]
    \centering
    \includegraphics[width=0.47\linewidth]{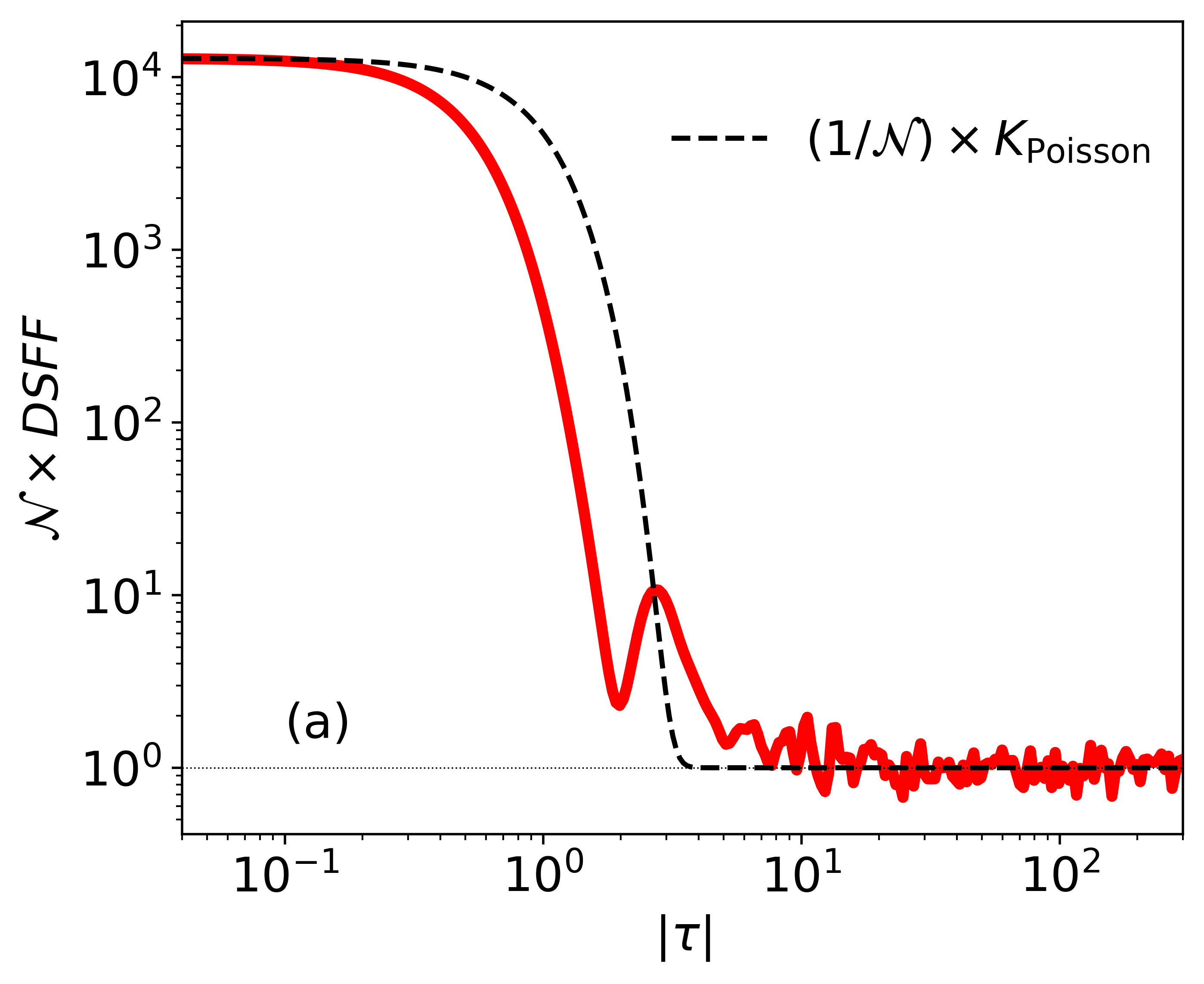}
    \includegraphics[width=0.47\linewidth]{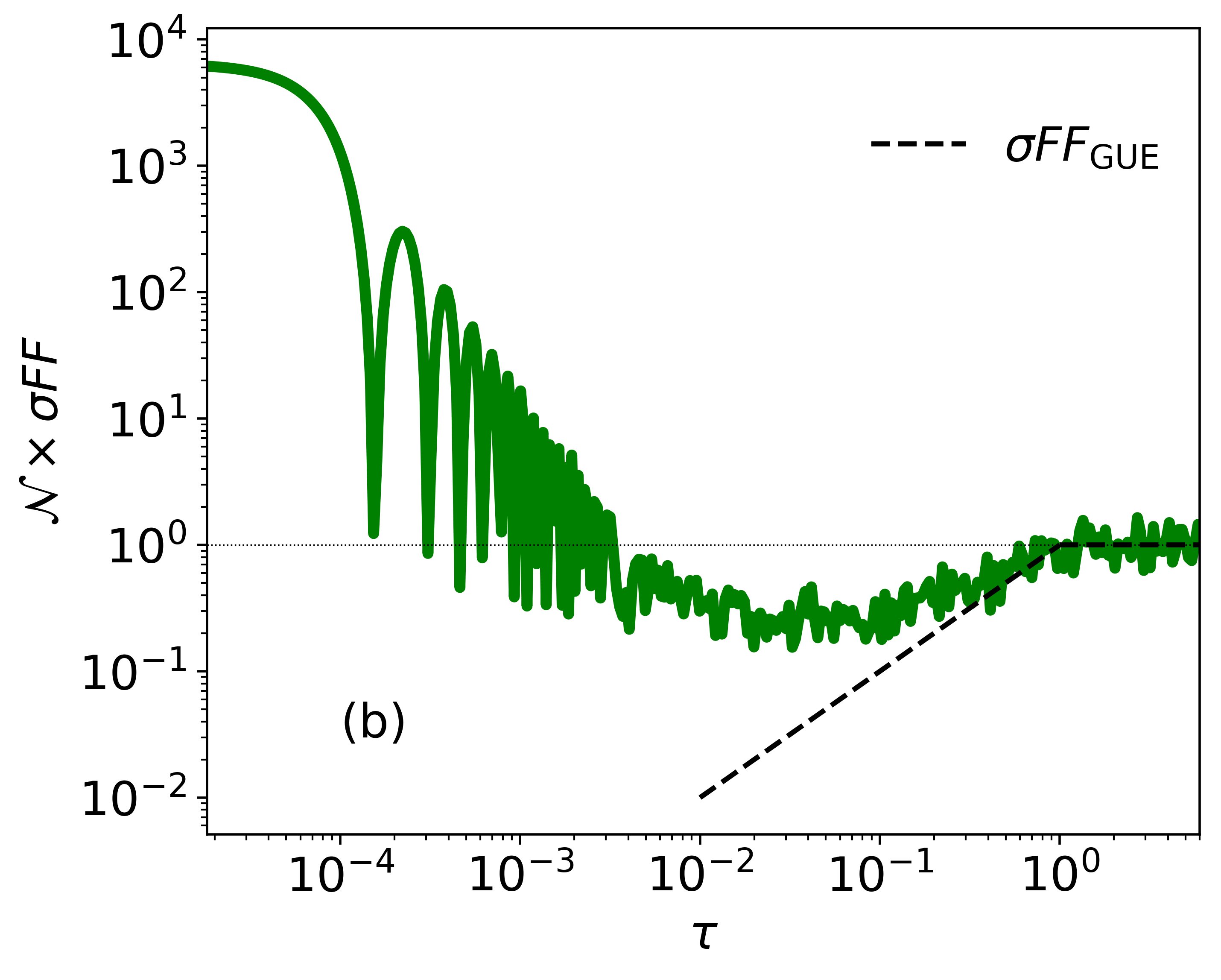}
    \caption{Plots for the (a) dissipative spectral form factor, given in Eq.~\eqref{DSFF} and (b) singular form factor, given in Eq.~\eqref{eq:sigma_FF} for the Hamiltonian model given in Eq.~\eqref{eq:model2} for the disorder strength $h=4.03$. The data is computed for system size $L=16$ and averaged over 20 realizations.}
    \label{fig:model2_form}
\end{figure}

\begin{figure}[h]
    \centering
    \includegraphics[width=0.47\linewidth]{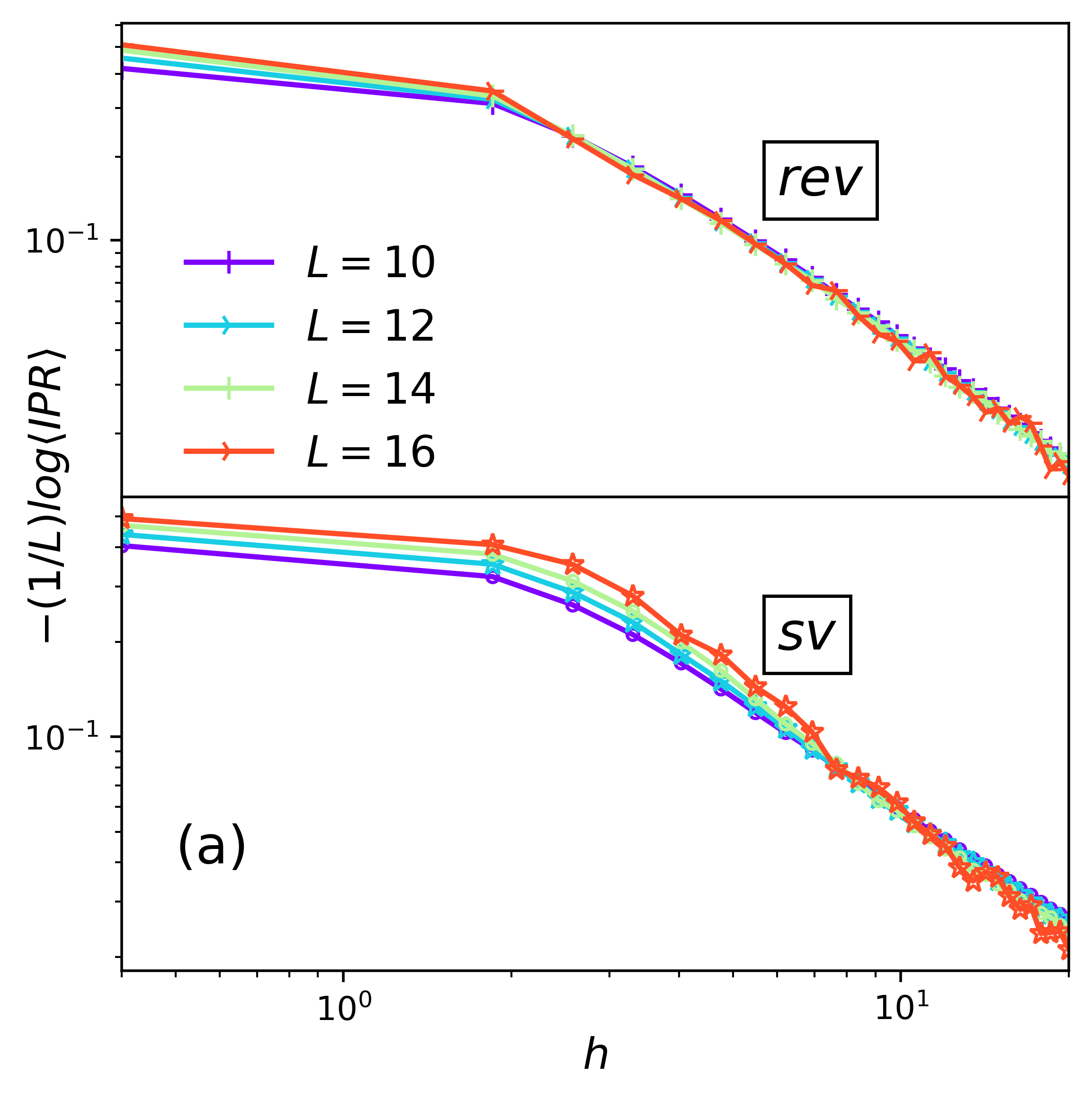}
    \includegraphics[width=0.47\linewidth]{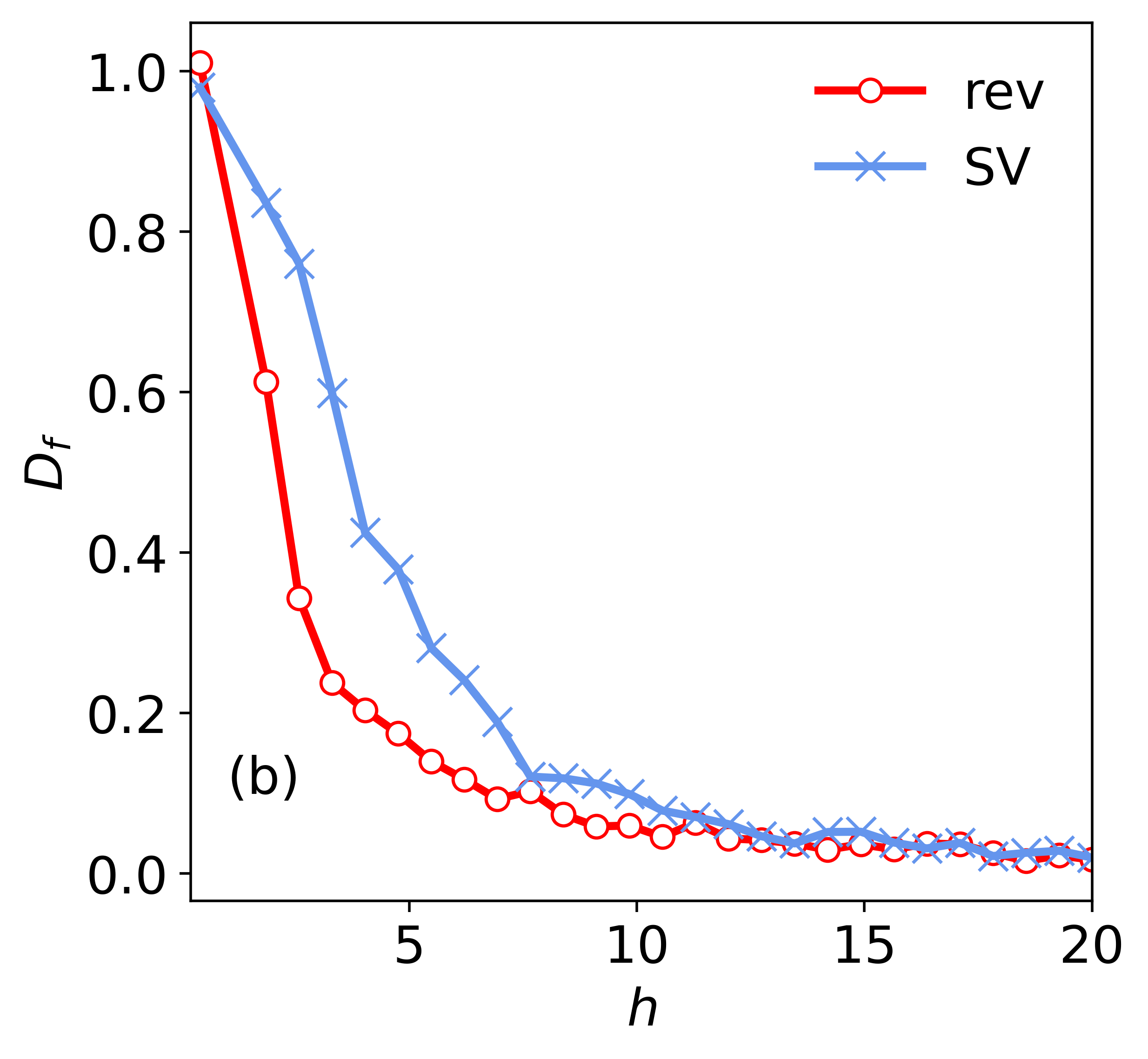}
    \caption{Plots for the (a) inverse participation ratio ($\rm IPR$), defined in Eq.~\eqref{IPR-frac} for $q=2$, constructed following the right eigenvectors (rev) of the complex eigenvalues $\langle \rm IPR \rangle_{\rm rev}$ and the IPR calculated using the singular vectors (sv) $\langle \rm IPR \rangle_{\rm sv}$ and (b) the fractal dimension $D_f$ for $q=2$ of the model given in Eq.~\eqref{eq:model2}, as a function of disorder strength $h$. The data is averaged over $12000$, $4000$, $1000$ and $20$ for the system size $L=10$, $12$, $14$, $16$ respectively.} 
    \label{fig:model2_IPR}
\end{figure}

{\it Non-Hermitian Hamiltonian with complex on-site disorder.--}  We next study a class of physically motivated NHHs whose Hermitian counterpart hosts different phases. Once again, we assess the utility of singular values compared to the analysis of complex spectra. The Hamiltonian that we investigate here reads as
\begin{equation}
  \label{eq:model2}
    H\!=\!\sum_{j=1}^{L}\Big[-t \big (\hat{c}^{\dagger}_j \hat{c}_{j+1}\!+\!{\rm h.c.} \big)+ (h_{{\rm Re},j}+i\,h_{{\rm Im},j}) \hat{n}_j+V \hat{n}_j \hat{n}_{j+1}\Big],
\end{equation}
where $\hat{c}_j$ ($\hat{c}_j^{\dagger})$ is the annihilation (creation) operator for fermions at the site $j$, $\hat{n}_j=\hat{c}_j^{\dagger} \hat{c}_j$ is the number operator at the $j$-th site, and $h_{{\rm Re}, j} \in [-h, h]$ and $h_{{\rm Im}, j} \in [0, h]$ are real and imaginary components of the random onsite disorder which are drawn independently from a uniform distribution for each site $j$, separately. For our analysis, we take $t=1$ and $V=2$. In Fig.~\ref{fig:model2_r}(a) and (b), we plot $\langle r \rangle_{\rm CSR}$ and $\langle r \rangle_{\rm SV}$ as a function of disorder strength $h$. We see that different transition points are manifested following the two different spectral analysis. While $\langle r \rangle_{\rm CSR}$ predicts a transition from ergodic to localized phase at $h \approx 3$, the $\langle r \rangle_{\rm SV}$ indicates a transition at $h \approx 7$. This estimate arising from singular value analysis is seemingly close to that of its Hermitian counterpart [i.e., setting $h_{{\rm Im}, j}=0$ in Eq.~\eqref{eq:model2}] (see supplemental material \cite{supp}). This is reminiscent of the observation made for the NH-PLBRM models in terms of the seemingly non-trivial resemblance between the singular value predictions and the predictions on the Hermitian counterpart of the underlying non-Hermitian random matrix. We further plot the dissipative spectral form factor and the singular form factor in Fig.~\eqref{fig:model2_form}(a) and (b), for a representative disorder strength $h=4.03$. We notice that the ramp observed in the singular form factor is completely missing in the DSFF. Therefore, our findings reveal that singular values that contain less information than the actual complex eigenvalues can erroneously suggest that the system is in the chaotic phase, whilst the complex eigenvalues predict otherwise. Finally, for the model given in Eq.~\eqref{eq:model2}, we present results for the IPR and the fractal dimension, defined in Eq.~\eqref{IPR-frac}. As evident in Fig.~\ref{fig:model2_IPR}, the different transition points are clearly reflected through these diagnostics when computed using the right eigenvectors and singular vectors.

\begin{figure}
    \centering
    \includegraphics[width=0.47\linewidth]{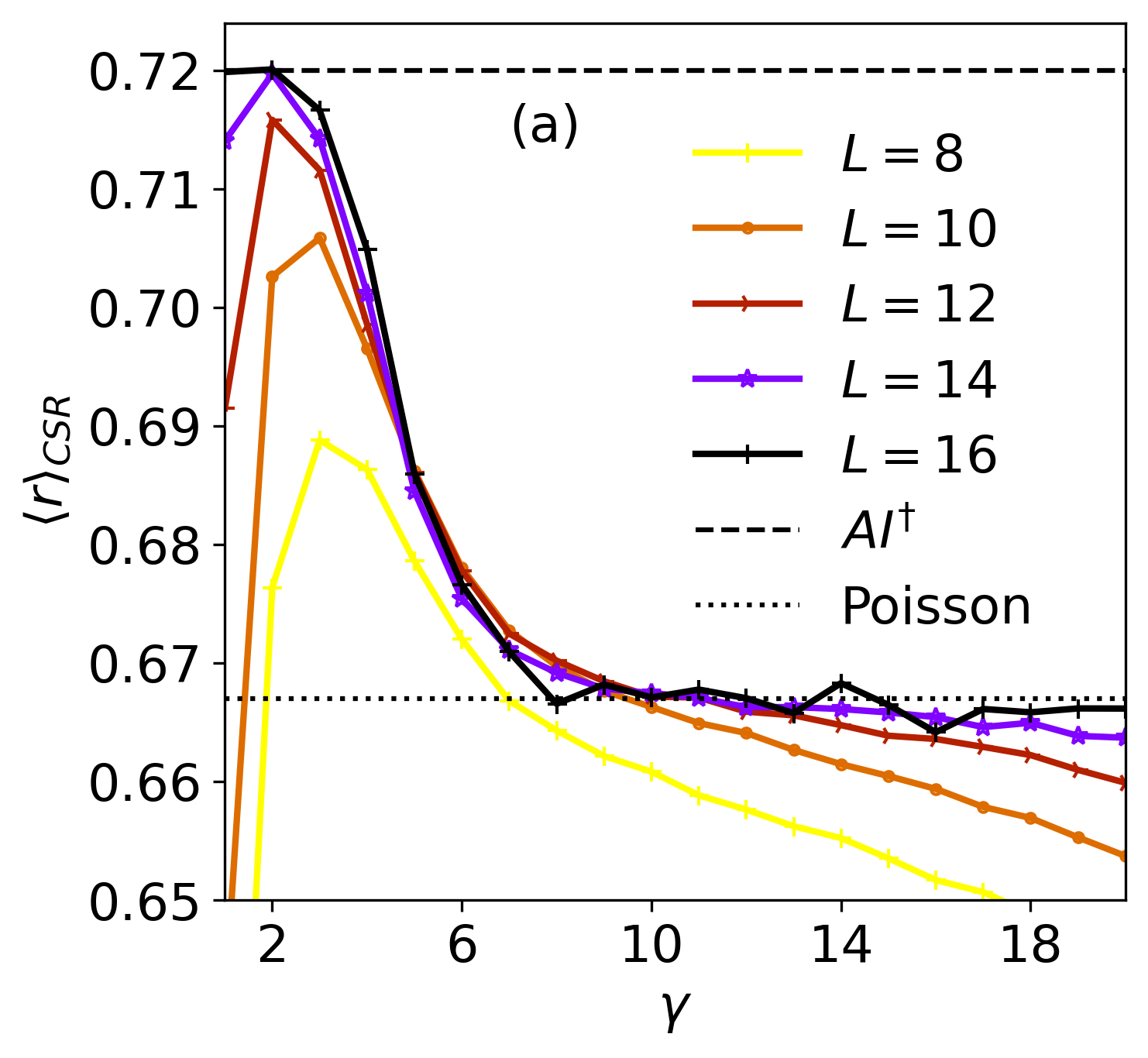}
    \includegraphics[width=0.47\linewidth]{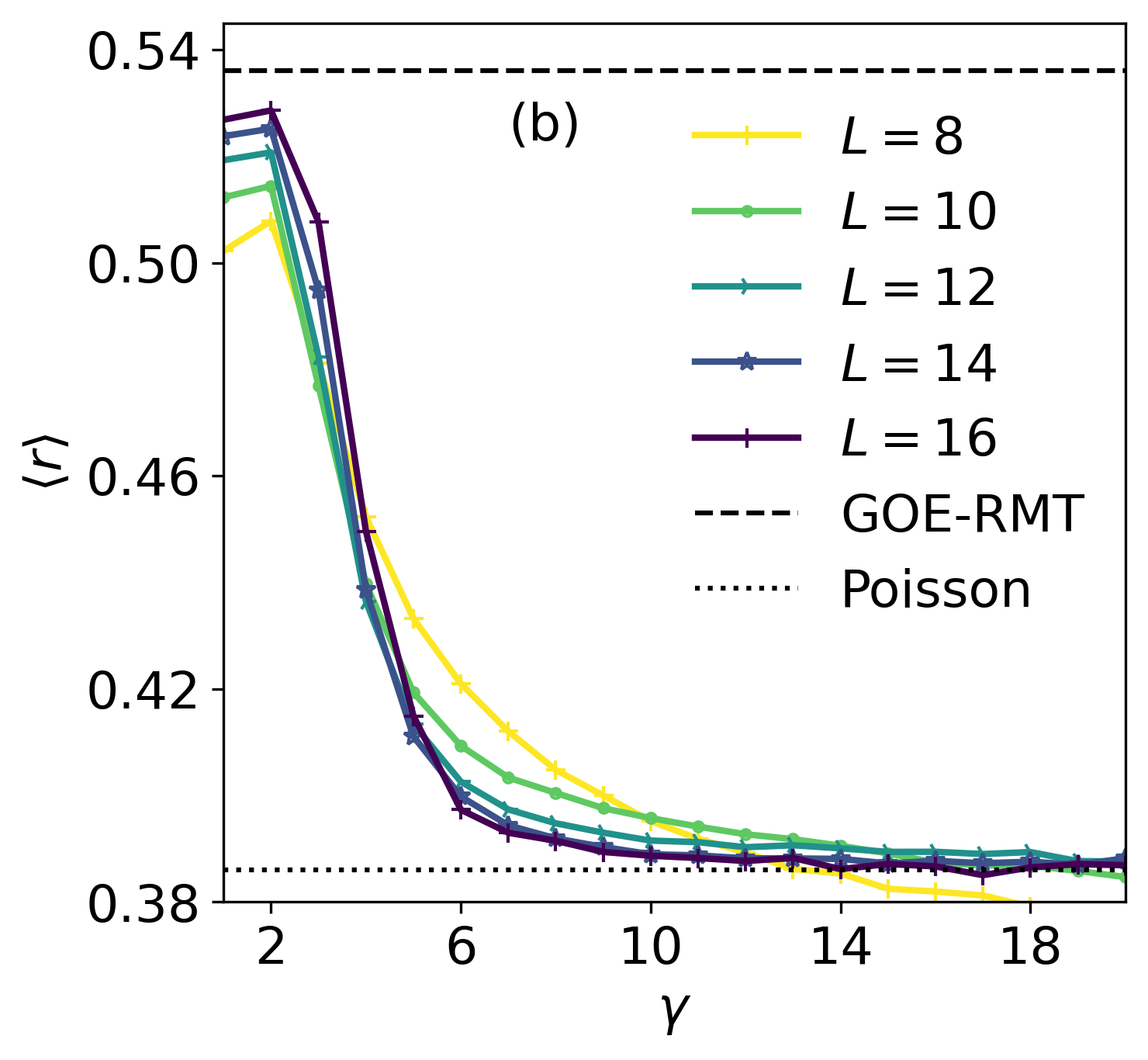}
    \caption{Plots for the average adjacent gap ratio statistics of (a) complex spectra $\langle r \rangle_{\rm CSR}$ and, (b) singular values $\langle r \rangle _{\rm SV}$ of the NHH given in Eq.~\eqref{xxz-nonherm} as a function of $\gamma$ for different system sizes. The top and bottom horizontal lines in (a) and (b) denote the expected values for the corresponding random matrix classes and the Poisson distributions, respectively. The data is averaged over $40000$, $12000$, $4000$, $1000$ and $20$ sample for the system size $L=8$, $10$ $12$, $14$, $16$, respectively.}
    %For this model, the transition between ergodic to localization is caused purely by the non-Hermitian disorder. Interestingly, for such class, the predictions for the transition point following the complex spectra and the singular values are almost identical.}
    \label{fig:model3_r_main}
\end{figure}

So far, we have discussed scenarios in which singular values predict seemingly different behavior as compared to the predictions via the complex spectra analysis. One, therefore, needs caution while identifying different phases using singular values/vectors. As a next step, one may naturally wonder for what kind of platforms singular value analysis provides a reasonably good estimate of different phases such as the ones provided by complex spectra. To differentiate from the previously discussed NH-PLBRM models and NHH models with complex on-site disorder, we now consider here a completely different class where the transition is purely induced by non-Hermitian disorder. The Hamiltonian we consider is a spin-chain given by \cite{chenu_svd}
\begin{equation}
H\!=\! \sum_{j=1}^{L} J \Big(\hat{S}_j^{x} \hat{S}_{j+1}^{x} + \hat{S}_j^{y} \hat{S}_{j+1}^{y} + \Delta \hat{S}_j^{z} \hat{S}_{j+1}^{z}\Big) - \frac{i \gamma_j}{2} \big(\hat{S}_j^{z} + \frac{1}{2}\big)\, ,
\label{xxz-nonherm}
\end{equation}
where $\gamma_j$ is the damping rate for the $j$-th site and each $\gamma_j$ is chosen independently from a uniform distribution over the interval $[0, \gamma]$. Note that in the absence of $\gamma$, the system is Hermitian and becomes the quantum integrable XXZ spin chain. Therefore, any transition in this model is solely caused by the non-Hermitian term that involves $\gamma$ in Eq.~\eqref{xxz-nonherm}. In fact, via the diagnostics involving singular values, this model is known to possess a transition from ergodic to localized phase as one increases $\gamma$ \cite{chenu_svd}. Here, we attempt to find out if such transitions and their locations are also predicted in complex spectral analysis. For such a class, we find strong evidence that analysis using singular value works remarkably well when compared with analysis with complex spectra. In Fig.~\ref{fig:model3_r_main} (a) and (b), we plot $\langle r \rangle_{\rm CSV}$ and $\langle r \rangle_{\rm SV}$ as a function of damping rate $\gamma$ for different system sizes and observe not only the transition in both singular value and complex value analysis but also the estimate of location is remarkably close ($\gamma \approx 6$). This analysis strongly hints that singular value analysis is powerful and justified in classes of models where transitions are solely produced by the non-Hermitian terms present in the Hamiltonian. 

{\it Summary.--} We thoroughly assess the utility of 
singular value in terms of its ability to characterize different spectral phases of non-Hermitian systems. More precisely, we find that the prediction for the transition lines/points following the singular value analysis seemingly differs from the one carried out using complex spectral analysis for a wide class of models. We demonstrate this using a non-Hermitian version of power-law banded random matrices, which is a paradigmatic model to understand transitions in non-Hermitian systems. Furthermore, we elucidate this result via a thorough analysis of a physical non-Hermitian many-body quantum Hamiltonian with complex onsite disorder. Finally we present scenarios where singular values can potentially be a very powerful tool in characterizing different phases. Our study reveals that  one such scenario is when transitions are solely caused by the non-Hermitian disorder term in the Hamiltonian.

Our work opens up further directions in understanding the information encoded and information unseen in singular values through the lens of different diagnostics for non-Hermitian matrices associated with non-Hermitian systems. Singular values, in several cases, can misjudge the location of the transition point. Nonetheless, it still contains rich information about the transition. As a consequence, it remains a promising avenue for characterizing non-Hermitian systems. It is an interesting and challenging question to isolate the precise information that is lost~\cite{baggioli2025} in singular values as compared to those contained in the underlying complex spectra.

{\it Acknowledgements.}-- M.K. acknowledges support from the Department of Atomic Energy, Government of India, under project No.~RTI4001. B.K.A acknowledges CRG Grant No.~CRG/2023/003377 from the Science and Engineering Research Board (SERB), Government of India. 
%B.K.A. would like to acknowledge funding from the National Mission on Interdisciplinary  Cyber-Physical  Systems (NM-ICPS)  of the Department of Science and Technology (DST), Govt.~of  India through the I-HUB  Quantum  Technology  Foundation, Pune, India. 
B.K.A thanks the hospitality of the International Centre of Theoretical Sciences (ICTS), Bangalore, India, under the associateship program. M.K. thanks the hospitality of the Department of Physics, IISER, Pune. M.K. thanks the hospitality of the Department of Physics, Princeton University, and Initiative for Theoretical Sciences, Graduate Center, City University of New York, USA.
\bibliography{references}

%\twocolumngrid

\clearpage
\setcounter{equation}{0}
\setcounter{figure}{0}
\renewcommand{\theequation}{S\arabic{equation}}
\renewcommand{\thefigure}{S\arabic{figure}}

\begin{center}
{\textbf{\underline{Supplementary Material}}}
\end{center}

%\section{Results}
\section{Additional results for non-Hermitian Hamiltonian with purely imaginary disorder}
\label{sec:ph}
\begin{figure}[h]
    \centering
    \includegraphics[width=0.47\linewidth]{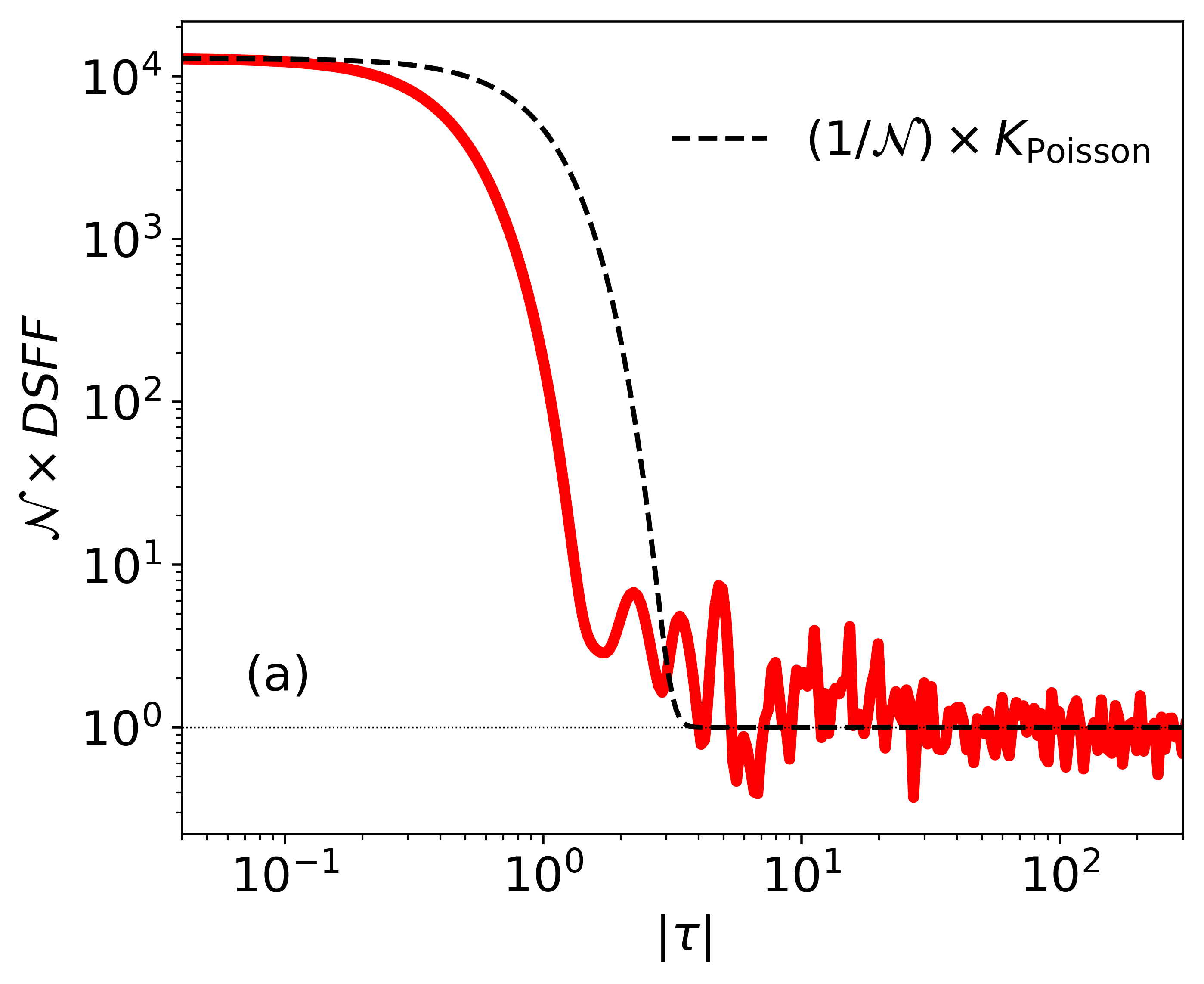}
    \includegraphics[width=0.47\linewidth]{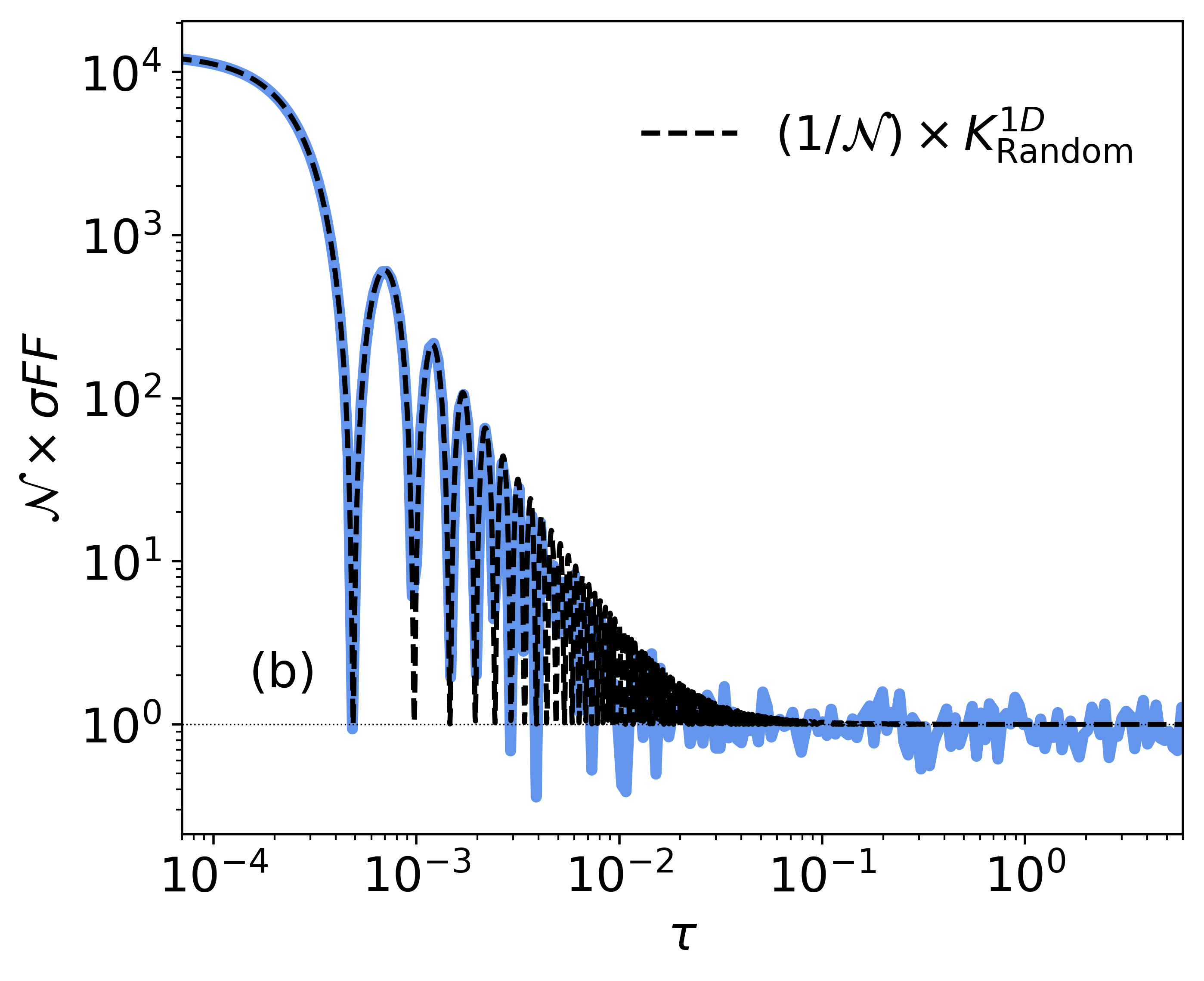}
    \caption{Plots for the (a) dissipative spectral form factor, given in Eq.~\eqref{DSFF} and (b) singular form factor, given in Eq.~\eqref{eq:sigma_FF}, for the Hamiltonian model given in Eq.~\eqref{xxz-nonherm_supp} for the disorder strength $\gamma=9.0$. The dashed line in (a) represents the analytical prediction that follows from the two-dimensional Poisson distribution ${K(\tau,\tau^{*})}_{\rm Poisson}={\cal N}+{\cal N}({\cal N}-1)\,e^{-|\tau|^2}$. The dashed line in (b) represents the analytical prediction~\cite{PhysRevResearch.3.L012019} that follows from one-dimensional uncorrelated random numbers given in Eq.~\eqref{eq:k1drandom}. The data is computed for system size $L=16$ and averaged over 20 realizations }
    \label{fig:xxz-nonherm_form}
\end{figure}

\begin{figure}
    \centering
    \includegraphics[width=0.8\linewidth]{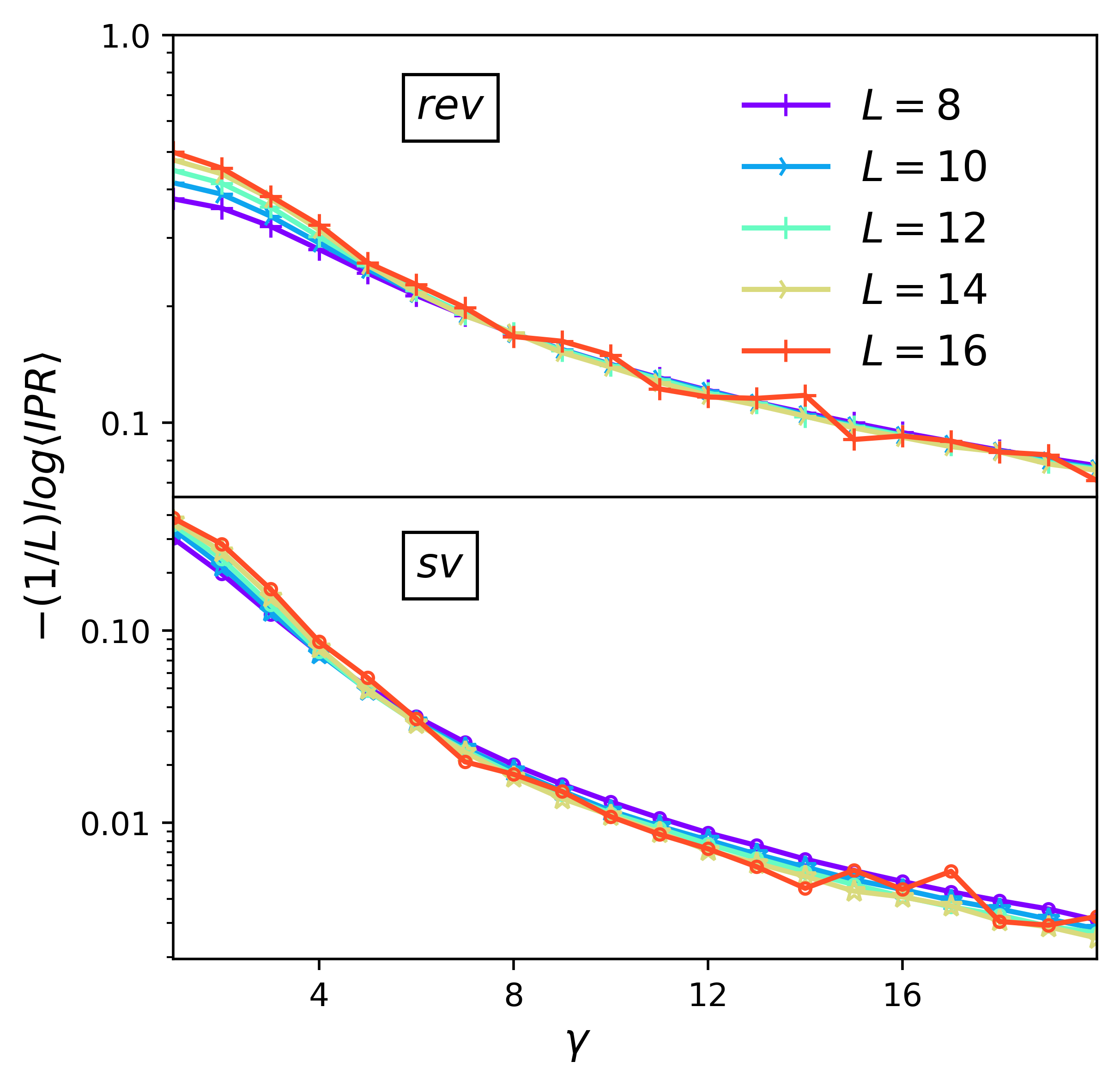}
    \caption{Plots for the inverse participation ratio ($\rm IPR$), defined in Eq.~\eqref{IPR-frac} for $q=2$, constructed following the right eigenvectors of the complex eigenvalues $\langle \rm IPR \rangle_{\rm rev}$ (top panel) and the IPR calculated using the singular vectors $\langle \rm IPR \rangle_{\rm sv}$ (bottom panel) of the model given in Eq.~\eqref{xxz-nonherm_supp} for different system sizes. The data is averaged over $40000$, $12000$, $4000$, $1000$ and $20$ realizations for the system size $L=8$, $10$ $12$, $14$, and $16$, respectively.}
    \label{fig:imag_disorder_IPR}
\end{figure}

In the section, we present additional results on the model where the transition is solely induced by non-Hermitian disorder. We recall that the  Hamiltonian of the spin-chain is given as \cite{chenu_svd}
\begin{eqnarray}
H &=& \sum_{j=1}^{L} J \Big(\hat{S}_j^{x} \hat{S}_{j+1}^{x} + \hat{S}_j^{y} \hat{S}_{j+1}^{y} + \Delta \hat{S}_j^{z} \hat{S}_{j+1}^{z}\Big)  \nonumber \\ &-&\frac{i \gamma_j}{2} \big(\hat{S}_j^{z} + \frac{1}{2}\big)\, , 
\label{xxz-nonherm_supp}
\end{eqnarray}
where $\gamma_j$ is the damping rate for the $j$-th site and each $\gamma_j$ is chosen independently from a uniform distribution over the interval $[0, \gamma]$. 
We plot the dissipative spectral form factor and the singular form factor in Fig.~\eqref{fig:xxz-nonherm_form}(a) and (b), respectively, for a representative point $\gamma=9.0$. In this case, we notice that the ramp is absent even in the singular form factor. This is unlike what we have reported in Fig.~\ref{fig:model2_form} where the Hamiltonian had complex on-site disorder [Eq.~\eqref{eq:model2}]. For the DSFF, the result resembles an analytical prediction that follows from the two-dimensional Poisson distribution ${K(\tau,\tau^{*})}_{\rm Poisson}={\cal N}+{\cal N}({\cal N}-1)\,e^{-|\tau|^2}$. For the $\sigma FF$, the result matches well with the analytical prediction that follows from one-dimensional uncorrelated random numbers given as \cite{PhysRevResearch.3.L012019}
%We use analytical expression of spectral for factor for completely uncorrelated levels~\cite{PhysRevResearch.3.L012019} to compare our data of $\sigma FF$ in localized phase that is given by,
\begin{equation}
    K_{\rm random}^{1D}(\tau)= {\cal N} + {\cal N} ( {\cal N } -1) \Big | \frac{\sin(\mu{\cal N} \tau /2) }{(\mu{\cal N} \tau /2)} \Big |^{2}\, ,
    \label{eq:k1drandom}
\end{equation}
where $\mu$ is the mean level spacing which is $1$ for unfolded spectra. In Fig.~\eqref{fig:imag_disorder_IPR}(a) and (b), we present results for the IPR, defined in Eq.~\eqref{IPR-frac} of the main text, as a function of damping rate $\gamma$. As evident in  Fig.~\eqref{fig:imag_disorder_IPR}, we find similarity in the location of transition points when computed using the right eigenvectors and singular vectors. This section therefore highlights the power of information encoded in singular values when dealing with non-Hermitian systems where transitions are solely driven by imaginary disorder. 

\section{Result for the Hermitian limit of the Hamiltonian given in Eq.~(\ref{eq:model2}) of the main text}
In this section, we provide result for adjacent gap ratio as a function of disorder strength $h$ for the Hermitian version of the model defined in Eq.~\eqref{eq:model2} of the main text, i.e., we set $h_{{\rm Im}, j}=0$ which gives the following Hermitian Hamiltonian
\begin{equation}
  \label{eq:model2-Hermitian-supp}
    H\!=\!\sum_{j=1}^{L}\Big[-t \big (\hat{c}^{\dagger}_j \hat{c}_{j+1}+ {\rm h.c.} \big)+ h_j \, \hat{n}_j+V \hat{n}_j \hat{n}_{j+1}\Big].
\end{equation}
Recall that for the non-Hermitian model, following the $\langle r \rangle_{\rm CSR}$ and $\langle r \rangle_{\rm SV}$ the transition from ergodic to localized phase was observed at $h\approx 3$ and $h \approx 7$, respectively. The Hermitian counterpart of this model shows transition around $h \approx 8$ which is seemingly close to that predicted via singular values. 

\begin{figure}
    \centering
    \includegraphics[width=0.57\linewidth]{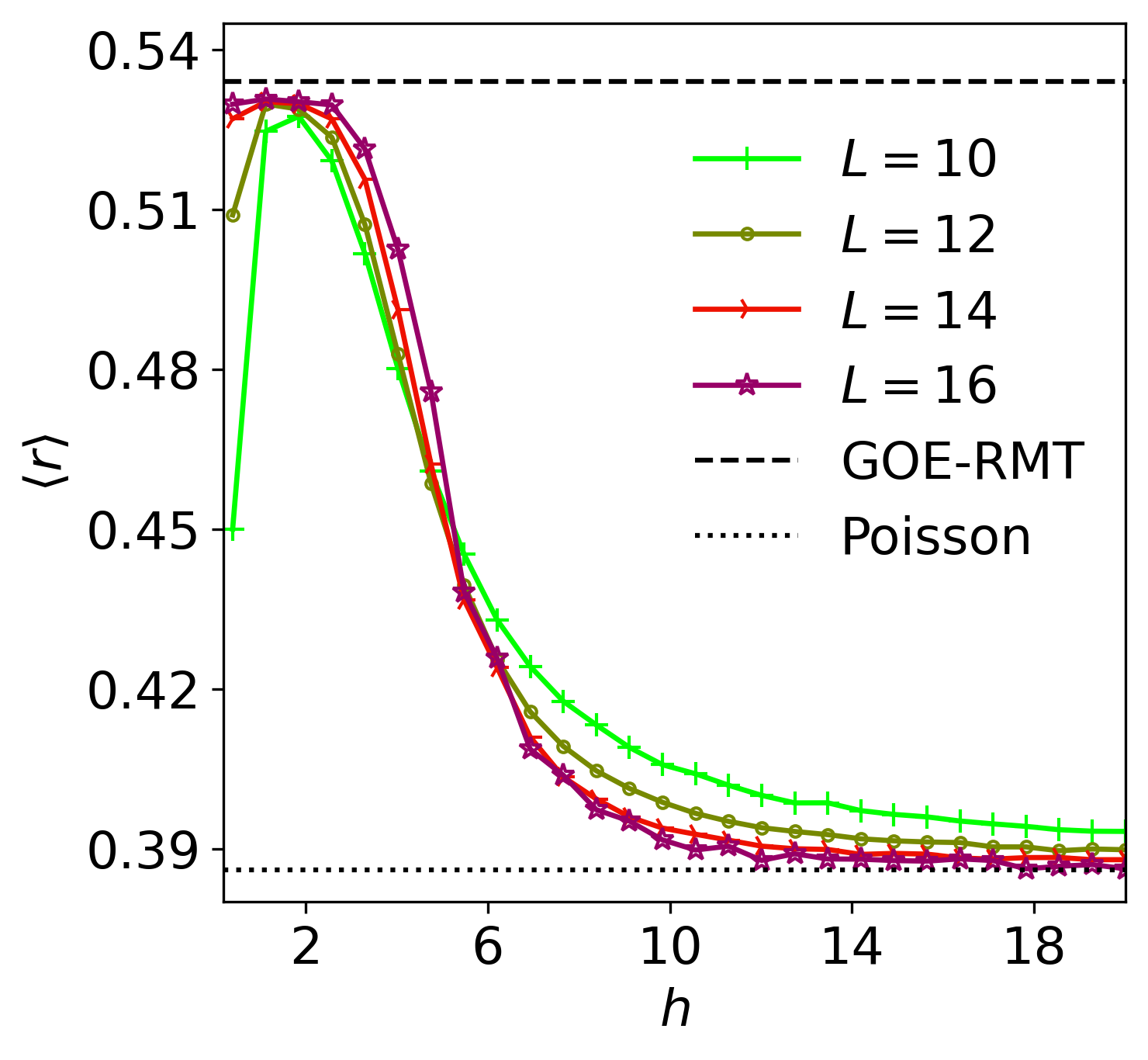}
    \caption{Plot for the average gap ratio statistics of the Hermitian limit of the model defined in Eq.~\eqref{eq:model2-Hermitian-supp}. The plot indicates a transition from ergodic to localized phase around $h \approx 8$. This value is close to the estimate that is obtained via singular values of the non-Hermitian version of the model, given in Eq.~\eqref{eq:model2}. The top dashed line indicates the prediction that follows from Gaussian orthogonal ensemble (GOE)-RMT and the bottom dotted line indicates that of the Poisson statistics. The data is averaged over $12000$, $4000$, $1000$ and $20$ samples for the system size $L=10$, $12$, $14$, and $16$, respectively.}
    \label{fig:model3_r}
\end{figure}
\label{sec:pherm}

\end{document}